\documentclass[12pt]{iopart}

\usepackage[cp866]{inputenc}
\usepackage[english]{babel}
\usepackage{graphicx}
\usepackage{bm}
\usepackage{epsf}
\usepackage{iopams}

\begin{document}

\title[Electron shear viscosity]
{Shear viscosity of degenerate electron matter}

\author{P.~S.~Shternin}
\address{
        Ioffe Physical Technical Institute,
        Politekhnicheskaya 26, 194021 Saint-Petersburg, Russia}
\ead{pshternin@gmail.com}

\begin{abstract}
We calculate the partial electron shear viscosity
$\eta_{ee}$ limited by
electron-electron collisions in a strongly
degenerate electron gas taking into account the Landau damping of
transverse plasmons. The Landau damping strongly suppresses
$\eta_{ee}$ in the domain of ultrarelativistic
degenerate electrons and modifies its 
temperature behavior. The efficiency of the electron shear
viscosity in the cores of white dwarfs and envelopes of neutron
stars is analyzed.
\end{abstract}

\submitto{\JPA}
\maketitle

\section{Introduction}
\label{introduc}

Transport properties of degenerate dense matter have been
studied
for a long time. They are especially important for
simulating various processes in
neutron stars, white dwarfs and degenerate cores of giant stars.
All these objects contain a degenerate electron gas, and the
electrons can give considerable contribution to transport
coefficients. Although the electron transport problem is very well
elaborated and described in textbooks (e.g. \cite{ziman60}), some
aspects have to be reconsidered.

The first studies of electron transport properties (thermal
conductivity) of degenerate stellar matter were performed by
Marshak \cite{marshak41}, Mestel \cite{mestel50} and Lee
\cite{lee50} in the 1940s. Further work in the next two decades is
described by Lampe \cite{lampe68} who made considerable
contribution into the calculation of the electron thermal
conductivity. Later the subject has been studied by many authors,
particularly, by Flowers and Itoh \cite{fi76} and others
\cite{timmes92,uy80,gyp01,yash91,gy95}. However, all these authors
have considered collisions of relativistic charged particles in
the static limit. The importance of dynamical interactions for
relativistic particles was pointed out by Heiselberg and Pethick
\cite{hp93} who analyzed transport properties of an
ultrarelativistic quark plasma. The authors calculated transport
coefficients of a degenerate ultrarelativistic plasma and showed
that the results are qualitatively different from standard
Fermi-liquid expressions because ultrarelativistic charged
particles interact mainly through their currents. Such an
interaction is produced via the exchange of transverse plasmons
(instead of the standard Coulomb charge-charge interaction of
non-relativistic particles via the exchange of longitudinal
plasmons). An inclusion of transverse plasmons increases the
effective collision rates (decreases the kinetic coefficients).

Recently, we have reconsidered \cite{sy06} the electron thermal
conductivity of degenerate matter containing an electron gas of
any degree of relativity. We have shown, that a correct treatment
of electron-electron collisions, including the exchange of
transverse plasmons, increases the contribution of these
collisions into the electron thermal conductivity.

In the present paper we perform similar revision of the electron
shear viscosity. Such a viscosity is important for studying
hydrodynamical processes in neutron stars and white dwarfs,
particularly the damping of oscillations of these objects. We will
focus on the contribution from the electron-electron collisions.
Our analysis will be similar to that in Refs. \cite{hp93,sy06}.
Thus, we omit technical details.

The electron shear viscosity can
be written as \cite{ziman60}
\begin{eqnarray}
  \eta_{e}&=&\frac{n_e v_{e} p_{e}}{5\nu_{e}},
\label{eta}\\
\nu_{e}&=&\nu_{ee}+\nu_{ei},
\end{eqnarray}
where $n_e$ is the electron number density; $p_{e}=\hbar
(3\pi^2n_e)^{1/3}$ and $v_{e}=p_e/m_e^*$ are, respectively, the
electron Fermi-momentum and Fermi-velocity, $m_e^*$ being the
electron effective mass on the Fermi surface (it differs from bare
electron mass due to relativistic effects). Furthermore, $\nu_e$
is the total electron effective collision frequency, a sum of the
electron-ion collision frequency $\nu_{ei}$ and the
electron-electron collision frequency $\nu_{ee}$. Detailed
calculations of $\nu_{ei}$ have been recently performed in Ref.\
\cite{chy05}, while $\nu_{ee}$ is the main subject of our study.
After calculating $\eta_{ee}$, we briefly analyze the efficiency
of $\eta_e$ in the cores of white dwarfs and envelopes of neutron
stars (particularly, for the damping of pulsations of pre-white
dwarf and white dwarf stars).

\section{Formalism}
\label{form}

We consider an almost ideal and uniform
strongly degenerate electron gas. The electrons
can have any degree of relativity and
collide among themselves and with plasma
ions. To calculate the electron-electron collision frequency we use
the standard variational approach with the simplest trial function
\cite{ziman60}.
The variational expression for the effective electron-electron
collision frequency, that determines the shear viscosity, is
\begin{eqnarray}
\nu_{ee}&=& \frac{15\pi^2 \hbar^3}{8v_e p_e^4 k_B T}\int
\frac{{\rm d} \mathbf{p_1}{\rm d} \mathbf{p_2}{\rm d}
\mathbf{p_1'}{\rm d}
\mathbf{p_2'}}{(2\pi\hbar)^{12}}\nonumber\\
&\times& W(12|1'2') f_1 f_2 (1-f_1')(1-f_2')\nonumber\\
&\times&
\left[p_{1x}v_{1y}+p_{2x}v_{2y}-p_{1'x}v_{1'y}-p_{2'x}v_{2'y}\right]^2,
\label{general}
\end{eqnarray}
where $T$ is the temperature and $k_B$ is the Boltzmann constant.
The integration is performed over all possible electron states
involved into collisions $\mathbf{p}_1\mathbf{p}_2\to
\mathbf{p}_1'\mathbf{p}_2'$; $\mathbf{p}$ is an electron momentum,
$p_{x}$ being its component along the $x$ axis, $v_y$ is the
electron velocity component along the $y$ axis; primes indicate
particle states after a collision; $f$ is the Fermi-Dirac distribution
function. Equation (\ref{general}) includes the symmetry factor $1/2$
which excludes double counting of the same collision
events of identical particles (electrons). Furthermore,
$W(12|1'2')$ is the differential transition
probability, summed over spin states of colliding particles,
\begin{eqnarray}
  W(12|1'2')&=&\frac{(2\pi\hbar)^4}{\hbar^2}
  \delta(\varepsilon_1'+\varepsilon_2'-\varepsilon_1-\varepsilon_2)
\nonumber\\
  &\times&
  \delta(\mathbf{p}_1'+\mathbf{p}_2'-\mathbf{p}_1-\mathbf{p}_2)\sum_{spins}
  |M_{fi}|^2,
\end{eqnarray}
where $\varepsilon$ is the electron energy. The delta-functions reflect
momentum and energy conservation; $|M_{fi}|^2$ is the squared
matrix element.

The matrix element $M_{fi}$ for a collision of charged particles
depends on the character of plasma
screening of electromagnetic interaction
between these particles.
For $ee$ collisions (of identical particles), one has
$M_{fi}=M_{fi}^{(1)}-M_{fi}^{(2)}$, where $M_{fi}^{(1)}$
and $M_{fi}^{(2)}$ correspond to two collision channels, $1\to 1';\;2\to 2'$
and $1\to 2';\;2\to 1'$, respectively, and
\begin{equation}
M_{fi}^{(1)}=\frac{4\pi
e^2}{c^2}\left(\frac{J_{1'1}^{(0)}J_{2'2}^{(0)}}{q^2+\Pi_l}-
\frac{\mathbf{J}_{t1'1}\cdot\mathbf{J}_{t2'2}}{q^2-\omega^2/c^2+\Pi_t}\right).
\label{Mfi}
\end{equation}
In this case, $\hbar \mathbf{q}=\mathbf{p}_1'-\mathbf{p}_1$ and
$\hbar \omega=\varepsilon_1'-\varepsilon_1$ are momentum and
energy transfers in a collision event, respectively;
$J_{e'e}^{(\nu)}=(J_{e'e}^{(0)},
\mathbf{J}_{e'e})=(2m_e^*c)^{-1}(\bar{u}_{e'}\gamma^\nu u_e)$ is
the transition 4-current, $\mathbf{J}_{te'e}$ is the component of
$\mathbf{J}_{e'e}$ transverse to $\mathbf{q}$; $\gamma^\nu$ is a
Dirac matrix; $u_e$ a normalized electron bispinor (with
$\bar{u}_eu_e=2m_ec^2$), and $\bar{u}_e$ is a Dirac conjugate. The
expression for $M_{fi}^{(2)}$ is obtained from Eq.\
(\ref{Mfi}) by the interchange of indices $1'\leftrightarrow 2'$
(which results also in changing $\mathbf{q}\to \mathbf{q}$
and $\omega\to-\omega$).

For collisions of strongly degenerate particles, calculations are
simplified by placing all interacting particles on their Fermi
surfaces (whenever possible). Characteristic values of $q$ and
$\omega$ in electron-electron collisions are determined by plasma
screening, that is described by polarization functions $\Pi_l$ and
$\Pi_t$ for the longitudinal (charge-charge) and transverse
(current-current) interactions, respectively. The nature of plasma
screening is discussed in Refs.\ \cite{hp93,sy06}. Let us
summarize the main points. Usually, collisions of charged particles
are studied in the so-called
weak-screening approximation, in which
momentum transfers are smaller than particle momenta.
This approximation is justified by the long-range nature
of electromagnetic interactions reflected
in a specific $q$-dependence of the matrix
element (\ref{Mfi}) (a well pronounced peak at small $q$).
As a result, only small values of
$q$ contribute to the integral (\ref{general}),
being determined by the character of plasma
screening \cite{hp93}. In the weak screening approximation,
it is sufficient to
consider the polarization functions in the classical limit ($\hbar
q \ll p_e$ and $\hbar \omega \ll v_ep_e$), in which they are given
by
\cite{abr84}
\begin{equation}
    \Pi_l = q_0^2 \, \chi_l(x), \qquad
    \Pi_t = (q_0 v_e/c)^2 \, \chi_t(x),
\label{polariz}
\end{equation}
where $x=\omega /(q v_e)$,
\begin{eqnarray}
     \chi_l(x) &=&  1- {x \over 2}\,
    \ln \left( x+1 \over x-1 \right) ,
\nonumber \\
    \chi_t(x) &=&  {x^2 \over 2} +
    { x(1-x^2) \over 4}\, \ln \left( x+1 \over x-1 \right) ,
\label{chi}
\end{eqnarray}
$\hbar^2q^2_0=4e^2p_e^2/(\pi \hbar v_e)$, and $q_0$ is the
Thomas-Fermi electron screening wavenumber. The longitudinal and
transverse screenings are essentially different. The most striking
difference occurs in a strongly degenerate plasma at temperatures
$T$ much below the electron plasma temperature $T_{pe}=\hbar
\omega_{pe}/k_{\rm B}$ (that is determined by the electron plasma
frequency $\omega_e=\sqrt{4\pi e^2 n_e/m_e^*}$). In this case, it
is sufficient to consider the low-frequency limit of $\omega\to 0$
and $\omega/q\ll v_e$, in which
\begin{equation}
   \chi_l = 1, \qquad
   \chi_t= i \,
   \pi \omega /(4q v_e).
\label{polariz1}
\end{equation}
The longitudinal polarization function is real in this limit. It
means, that longitudinal electromagnetic
interaction (via the exchange of longitudinal
plasmons) results in the standard Debye-like
screening with a characteristic momentum transfer $q_l\sim q_0$ [see
Eq.\ (\ref{Mfi})]. For the transverse interaction
(via the exchange of transverse plasmons), characteristic momentum
transfer is different, $q_t\sim (\omega q_0^2/v_e)^{1/3}$.
Moreover, the transverse polarization function is pure imaginary.
Accordingly, virtual transverse plasmons undergo
collisionless absorption via
the well-known Landau damping.

Comparing screening lengths in the low-energy limit, we see that
$q_t\ll q_l$. Thus, transverse interactions occur on larger
length-scales than longitudinal ones and are, therefore, more
frequent. In the previous works on the electron shear viscosity,
this difference between longitudinal and transverse interactions
have been neglected and  one set $\Pi_l=\Pi_t=q_0^2$. It is a good
approximation in the non-relativistic case, because the ratio of
transverse to longitudinal parts of the matrix element in Eq.\
(\ref{Mfi}) contains a relativistic factor
$(J_{te'e}/J^{(0)}_{e'e})^2\sim v_e^2/c^2$. Hence, transverse
interactions of non-relativistic particles are inefficient. In
contrast, the collisions of relativistic electrons via the
exchange of transverse plasmons are more important than the
collisions via the exchange of longitudinal plasmons (owing to
larger screening length). This effect was analyzed by Heiselberg
and Pethick \cite{hp93} for a gas of ultrarelativistic quarks.
Here, we consider the electron-electron collisions for any degree
of electron relativity, in analogy with the study of Ref.\
\cite{sy06} for the thermal conductivity.

Placing all colliding particles on their Fermi surfaces and
performing possible analytical integrations with the aid of
delta-functions, we finally obtain the following expressions for
the electron-electron collision frequency and associated partial
shear viscosity,
\begin{equation}
  \nu_{ee}=\frac{12\alpha^2}{\pi\hbar} k_B T \frac{c^2}{v_e^2}I_\eta(u,\theta),
  \qquad
  \eta_{ee}= \frac{\pi \hbar n_ep_e v_e^3}{60\alpha^2c^2 k_BTI_\eta(u,\theta)}.
\label{result}
\end{equation}
Here, $\alpha=e^2/\hbar c$ is the fine structure constant, and
\begin{eqnarray}
   I_\eta(u,\theta) &=&
       \int_0^\infty {\rm d}w \, { w \, {\rm e}^w
       \over ({\rm e}^w-1)^2 } \, \int_0^1 {\rm d}x\,(1-x^2)
       \int_0^{ \pi} {{\rm d} \phi \over  \pi}\,(1-\cos \phi)
\nonumber \\
  && \times  \left| {1 \over 1 + (x \theta /w)^2\,\chi_l(x)}-
      { u^2 (1-x^2) \,\cos \phi \over
      1- u^2 x^2 + u^2 (x \theta /w )^2\,\chi_t(x)} \right|^2
\label{I}
\end{eqnarray}
is a dimensionless function of two variables,
\begin{equation}
   u \equiv v_e/c, \qquad \theta= \hbar v_e q_0/(k_{\rm B}T)
   = \sqrt{3} T_{pe}/T;
\label{betatheta}
\end{equation}
$\phi$ is the angle between components of momenta
$\mathbf{p}_{1t}$ and $\mathbf{p}_{2t}$ transverse to
$\hbar\mathbf{q}$, and $w=\hbar\omega/(k_BT)$. Note, that in the
small-momentum approximation, the interference term between
$M^{(1)}_{fi}$ and $M^{(2)}_{fi}$ is small and both collision
channels give equal contributions, resulting in $|M_{fi}|^2\to
2|M_{fi}^{(1)}|^2$. Equations (\ref{result}) and (\ref{I})
generalize Eqs.\ (47) and (48) from Ref.\ \cite{hp93} to the case
of arbitrary degree of relativity.

Integration over  $\phi$ in Eq. (\ref{I}) is straightforward
and gives
\begin{equation}
I_\eta(u,\theta)=I_l(u,\theta)+I_t(u,\theta)+I_{tl}(u,\theta),
\end{equation}
where  $I_l$ comes from the exchange of
longitudinal plasmons, $I_t$ from the exchange of
transverse plasmons, and $I_{tl}$ is the interference term.

Following Ref.\ \cite{sy06}, consider four regimes of
collisions between degenerate electrons
 as indicated in Table \ref{tab:regimes}.

\begin{table}[b]
\caption[]{Four regimes of shear viscosity $\eta_{ee}$
of degenerate electrons.}
\label{tab:regimes}
\begin{center}
\begin{tabular}{c c c c c }
\hline \hline &~Electron~& &~~Main~~
& $T$-dependence \\
Regime&~velocity~&~Temperature~&~contribution~
& of $\eta_{ee}$ \\
\hline 
I  & $v_e \ll c$ &  $T \gtrsim T_{pe}$ & $I_l$  &
$[T\ln(T/T_{pe})]^{-1}$
  \\
II  & $v_e \ll c$ &  $T \ll T_{pe}$ & $I_l$  & $1/T^2$
  \\
III  & $v_e \approx c$ &  $T \gtrsim T_{pe}$ & $I_l+I_t+I_{lt}$ &
$[T\ln(T/T_{pe})]^{-1}$
  \\
IV  & $v_e \approx c$ &  $T \ll T_{pe}$ & $I_t$ & $1/T^{5/3}$
  \\
\hline \hline
\end{tabular}
\end{center}
\end{table}

The regime I occurs in a non-relativistic ($v_e\ll c$)
and rather warm (although degenerate) plasma ($T\gtrsim T_{pe}$).
The analysis of (\ref{I}) leads to the following
asymptotic expressions valid in this regime:
\begin{eqnarray}
   I_l&=& \frac{2}{3}\left(\ln
   \frac{1}{\theta}+1.919\right),
   \nonumber\\
I_t&=&
   \frac{8u^{4}}{35}\left(\ln\frac{1}{u\theta}+3.413\right),\\
   I_{tl}&=&\frac{8u^2}{15}\left(\ln
   \frac{1}{\theta}+2.512\right).
   \nonumber
\end{eqnarray}
The leading contribution comes from $I_l$ owing to a small
relativistic factor $u$ (as discussed above). The logarithmic
terms and constant corrections in brackets are corresponding
Coulomb logarithms (the corrections were calculated numerically).
The leading-term result for $\eta_{ee}$ in this regime is
well-known.

In the regime III, where the plasma is again warm
($\theta\gtrsim 1$), but the electron gas
is ultrarelativistic ($u \approx 1$),
$I_l$ is the same as
in the regime I (because $I_l$ is independent of $u$). The asymptotic
expressions for two other integrals have the same form as in
the regime I but with different corrections,
\begin{eqnarray}
  I_t&=& \frac{1}{3}\left(\ln\frac{1}{\theta}+2.742\right),
  \nonumber\\
  I_{tl}&=&\frac{2}{3}\left(\ln \frac{1}{\theta}+2.052\right).
\end{eqnarray}
Now all three terms give comparable contribution to
$I_\eta$. The asymptote for $I_\eta$ in the
ultrarelativistic case coincides with that
obtained by Heiselberg and Pethick
\cite{hp93}. It is
different from the asymptote obtained
by using an incorrect longitudinal
screening in the transverse part of
$M_{fi}$. However, the difference occurs only
in the corrections to the dominant logarithmic terms.

The regimes II and IV are realized in a cold plasma
($\theta\ll 1$). In this case, we get
\begin{eqnarray}
I_l&=&{\pi^3 \over 12 \theta},\nonumber\\
I_t&=&\xi\frac{u^{10/3}}{\theta^{2/3}},\qquad
\xi=\frac{\pi}{6}\left(\frac{4}{\pi}\right)^{1/3}
       \Gamma(8/3)\zeta(5/3)\approx 1.813,\nonumber\\
I_{tl}&=&\frac{\pi^3 u^2}{6\theta},
\end{eqnarray}
where $\zeta(z)$ is the Riemann zeta function and $\Gamma(z)$ is
the gamma function. In the ultrarelativistic
limit (the regime IV, $u=1$), these expressions coincide with those
obtained by Heiselberg and Pethick for an ultrarelativistic quark
plasma \cite{hp93}.

The main contribution to $I_\eta$ in the regime II comes from
$I_l$ due to a strong suppression of $I_{tl}$ and $I_{t}$ in the
non-relativistic case. In the regime IV, which operates usually
everywhere in the neutron star crust (excluding only a thin layer
-- a few meters -- from the surface) and in the neutron star core,
the situation is different. The leading term is then $I_t$, which
corresponds to the exchange of transverse plasmons. It dominates
because of smaller characteristic momentum transfers (larger
electron mean-free paths) than those owing to the exchange of
longitudinal plasmons. The temperature dependence of $\eta_{ee}$
in the regime IV differs from the standard Fermi-liquid case (now
$\eta_{ee} \propto T^{-5/3}$ instead of the standard dependence
$T^{-2}$). The difference is smaller, than for the thermal
conductivity $\kappa_{ee}$ ($\kappa_{ee}$ is remarkably
independent of $T$ in the regime IV while a standard Fermi-liquid
requires $\kappa_{ee} \propto T^{-1}$). Therefore, the Landau
damping affects the shear viscosity weaker than the thermal
conductivity.

In addition to the above asymptotes, we have calculated $I_l$,
$I_t$, and $I_{tl}$ on a dense grid of $u$ and $\theta$ values
covering transition regions between the regimes I--IV. We have
further obtained the fit expressions which reproduce numerical
results and asymptotic limits. The fit for $I_l$ is
\begin{equation}
I_l=\left(0.379+\frac{0.287}{1+0.165\theta
+0.0019\theta^2}\right)\ln \left(1+\frac{6.81}{\theta}\right),
\end{equation}
with the maximum fit error of $1.4\%$ at $\theta=1$ (let us
recall, that $I_l$ is independent of $u$).

The fit for $I_t$ reads
\begin{eqnarray}
 I_t&=&\left[ {1.813\over C}+
 \frac{(C_2-1.813/C)(1+534\theta u-764(\theta u)^2)}
 {1+705\theta u+1630(\theta u) ^2+(3+3.4u^2)(\theta u)^3}\right]
 \nonumber\\
&\times&\ln \left[1+\frac{C}{A(\theta u)^{1/3}+(\theta u)^{2/3}}\right],
\end{eqnarray}
where $C_1=0.7801+0.1337 u^4$, $C_2=0.686+0.315 u^4$,
$A=0.556-0.08u^2$, and $C=A\exp(C_1/C_2)$. The  maximum error of
$3.9 \%$ is at $\theta=10^{-4}$, $u=0.57$.

Finally, for $I_{tl}$ we obtain
\begin{eqnarray}
  I_{tl}&=&\left[{5.168 \over D}+\frac{(D_2-5.168/D)(1+p_1\theta
  -p_2\theta^2)}{1+p_3\theta+p_4\theta^2+p_5\theta^3}\right]
\nonumber\\
   &\times&\ln \left(1+\frac{D}{\theta}\right),
\end{eqnarray}
where $p_1=15.2-16.5u^2$, $p_2=0.354+1.178u^2$, $p_3=13.5-13.68
u^2$, $p_4=2.938-2.886u^2$, $p_5=0.00375-0.00373u^2$,
$D_1=1.339+0.029u^4$, $D_2=0.533+0.133u^4$, and $D=\exp(D_1/D_2)$.
The maximum error of $2.7 \%$ is at $\theta=10^{-2}$, $u=1$.

\section{Discussion}
\label{discuss}

Let us discuss the electron shear viscosity in a dense plasma. It is
convenient to rewrite Eq.~(\ref{eta}) as
\begin{eqnarray}
   \eta_e^{-1}=\eta_{ee}^{-1}+\eta_{ei}^{-1}, \quad
   \eta_{ee}=\frac{n_e v_e p_e}{5\nu_{ee}},\nonumber\\
   \eta_{ei}=\frac{n_e v_e p_e}{5\nu_{ei}},
\end{eqnarray}
where $\eta_{ee}$ and $\eta_{ei}$ are the partial
viscosities governed by electron-electron and electron-ion
collisions, respectively. We will discuss the effect of the Landau
damping on $\eta_{ee}$ and the contribution of $\eta_{ee}$
in the total electron shear viscosity $\eta_{e}$. The
partial viscosity $\eta_{ei}$ will be calculated
using the formalism of Ref.\ \cite{chy05}.

\begin{figure}
\begin{center}
\includegraphics[width=0.5\textwidth]{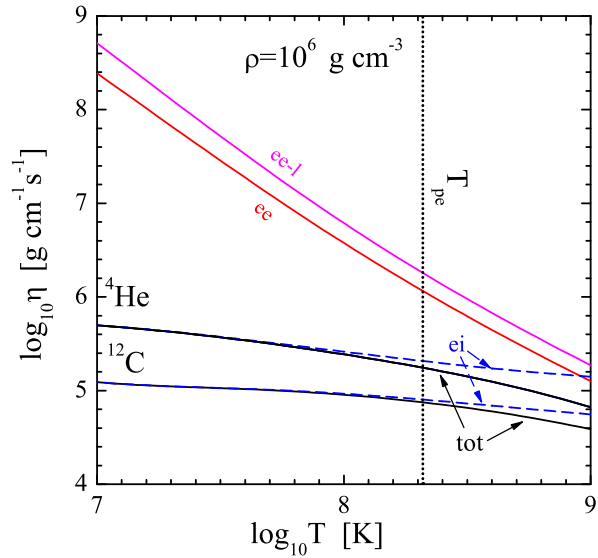}
\caption{Temperature dependence of the electron shear viscosity
at $\rho=10^6$~g~cm$^{-3}$. The line, marked
`ee', shows $\eta_{ee}$; the line `ee-l'
is the same, but only the contribution from the collisions via
the exchange of longitudinal plasmons is retained. The dashed
lines `ei' and the solid lines `tot' show $\eta_{ei}$ and
$\eta_e$, respectively, for helium and carbon plasmas.
Vertical dotted line indicates the electron plasma temperature.}
\label{F:rho6}
\end{center}
\end{figure}

Figure \ref{F:rho6} shows the temperature dependence of $\eta_e$
in the helium and carbon plasma at $\rho=10^6$~g~cm$^{-3}$ (where
degenerate electrons become mildly relativistic). At higher
$\rho$, the electrons are essentially relativistic and the effect
of the Landau damping is most pronounced, see Section \ref{form}.
Densities $\rho \gtrsim 10^6$ g~cm$^{-3}$ are appropriate for
degenerate cores of white dwarfs and red giants and for envelopes
of neutron stars. We plot the partial shear viscosity $\eta_{ee}$
 (line `ee') and the same viscosity, but retaining
only collisions via the exchange of longitudinal plasmons (line
`ee-l'). These lines are the same for the helium and carbon
plasmas. The relative contribution of collisions via the exchange
of transverse plasmons in $\nu_{ee}$ increases when $T$ falls
below the electron plasma temperature $T_{pe}$ ($\log_{10}
T_{pe}~[\rm K]\approx 8.32$, as indicated by the vertical dotted
line in Figure \ref{F:rho6}). The electron plasma temperature
separates the high-temperature and low-temperature asymptotic
regions (the regions III and IV for ultrarelativistic electrons, I
and II for non-relativistic electrons, see Table
\ref{tab:regimes}).

The dashed lines in Figure \ref{F:rho6} show
$\eta_{ei}$ and the solid lines marked `tot'
show the total electron viscosity $\eta_e$. One can see that
the electron-electron collisions are more efficient at higher temperatures
in a plasma containing lighter nuclei (more exactly,
the nuclei of lower charge, such as He, whose interaction
with electrons is especially weak).
At the given $\rho=10^6$ g~cm$^{-3}$,
the plasma screening type is not very
important, and the collisions via the exchange of transverse and
longitudinal plasmons give comparable contribution.
With increasing $\rho$,
the electron-electron collisions become
also more important at temperatures below $T_{pe}$.
However, at sufficiently high densities
low-charge nuclei transform to those with
higher charge (mainly owing to beta captures
and nuclear fusion reactions).
For these new nuclei with higher charge, the electron-electron collisions
become less important.

\begin{figure}
\begin{center}
\includegraphics[width=0.5\textwidth]{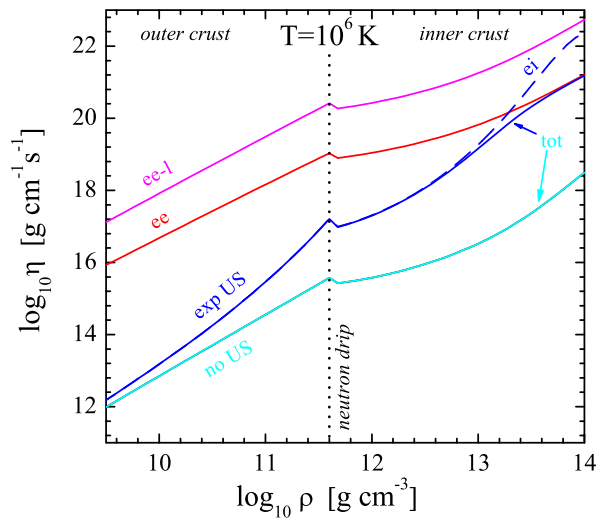}
\caption{ Density dependence of the electron shear viscosity in a
neutron star crust composed of the ground-state matter at
$T=10^6$~K. The line, marked `ee', shows $\eta_{ee}$; the line
`ee-l' is the same, but retaining the contribution from
longitudinal plasmons alone. The dashed line `ei'  displays solid
lines `tot' give $\eta_e$. The lines `no US' and `exp US' are
calculated neglecting the \emph{Umklapp} suppression and assuming
an exponential \emph{Umklapp} suppression in electron-ion
collisions, respectively. The vertical dotted line shows neutron
drip point. } \label{F:crust}
\end{center}
\end{figure}

At high densities in a neutron star envelope (crust) the
composition of matter cannot be arbitrary and is determined by an
evolution scenario for a given star. One usually considers the
models of ground-state (cold catalyzed) crust or accreted crust
(e.g., Ref.\ \cite{hpy07}). Figure \ref{F:crust} shows the density
dependence of $\eta_e$, $\eta_{ee}$ and $\eta_{ei}$ throughout a
neutron star crust (for $\rho>10^9$ g~cm$^{-3}$) at $T=10^6$~K. We
use a smooth composition model for the ground-state matter in the
crust \cite{hpy07}. Let us remind that the crust extends to $\rho
\sim 1.5 \cdot 10^{14}$ g~cm$^{-3}$, while higher $\rho$
correspond to a neutron star core composed of uniform neutron-rich
nuclear matter. The vertical dotted line on the Figure
\ref{F:crust} shows the neutron drip point
($\rho_{nd}=4.3\cdot10^{11}$~g~cm$^{-3}$), which separates the
outer crust (composed of electrons and nuclei) and the inner crust
(where free neutrons appear in dense matter). The lines marked
`ee' and `ee-l' again show $\eta_{ee}$ calculated including and
excluding the exchange of transverse plasmons, respectively.
Correct values of $\eta_{ee}$ are more than one order of magnitude
lower than the values `ee-l'. However, this suppression of
$\eta_{ee}$ by the Landau damping is insufficient for $\eta_{ee}$
to dominate in $\eta_e$. The solid line marked `no US' in Fig.\
\ref{F:crust} shows the total electron shear viscosity $\eta_e$ in
which $\eta_{ei}$ is calculated under the same assumptions as in
\cite{chy05}. Most importantly, the calculations neglect the
freezing of \emph{Umklapp} processes of electron-ion
(electron-phonon) scattering at low temperatures ($T \ll T_{pe}$)
due to band-structure effects associated with motion of the
electrons in crystalline lattice. It turns out that $\eta_{ei}$ is
several orders of magnitude lower than $\eta_{ee}$, and,
therefore, $\eta_e \approx \eta_{ei}$.

However, the freezing of \emph{Umklapp} electron-phonon scattering
processes is a delicate task which has not been studied in detail
in the literature. If the freezing operates, it enhances
$\eta_{ei}$ and makes $\eta_{ee}$ more important. To illustrate
this effect, in Fig.\ \ref{F:crust} we present the $\eta_e$ and
$\eta_{ei}$ curves, marked by `exp US' and calculated assuming an
exponential freezing of the electron-phonon scattering rate after
$T$ falls below some temperature $T_u$ (that is approximately two
orders of magnitude lower than ion plasma temperature). This
illustrative model was used in Ref.\ \cite{gyp01} (and was
described there in more detail). In this case the
electron-electron collisions become significant at the bottom of a
cold neutron star crust, at $\rho\gtrsim 10^{13}$~g~cm$^{-3}$.

\begin{figure}
\begin{center}
\includegraphics[width=0.5\textwidth]{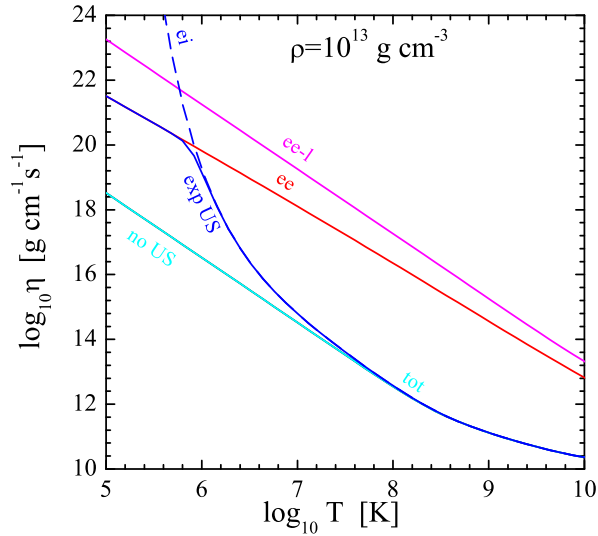}
\caption{Temperature dependence of the electron shear viscosity in
a neutron star crust composed of ground-state matter at
$\rho=10^{13}$~g~cm$^{-3}$. Notations are the same
as in Fig. \ref{F:crust}.}
\label{F:rho13}
\end{center}
\end{figure}


Figure \ref{F:rho13} shows the temperature dependence of $\eta_e$,
$\eta_{ei}$, and $\eta_{ee}$ at $\rho=10^{13}$ g~cm$^{-3}$ (deep
in the inner crust) under the same assumptions as in Fig.\
\ref{F:crust}. If the freezing of {\it Umklapp} processes is
neglected, then even for very low temperatures $T\sim
10^{5}$--$10^{6}$~K (where $\eta_{ee}$ is actually two orders of
magnitude lower than $\eta_{ee-l}$), $\eta_{ee}$ gives negligible
contribution to $\eta_e$ (Figure \ref{F:rho13}; curves `ee',
`ee-l' and 'tot', 'no US'). However, if one includes the freezing
of \emph{Umklapp} processes, then $\eta_{ee}$ dominates at low
$T$.

The suppression of \emph{Umklapp} processes at $T \lesssim T_u$
can actually be non-exponential, but power-law (A.~I.\ Chugunov,
private communication, 2007), which will reduce the importance of
the electron-electron collisions. This suppression is complicated
and should be a subject of separate study. In addition, we have
neglected scattering of electrons by charged impurities. It can
also be important at low temperatures (depending on charges and
abundance of impurity ions, e.g., \cite{pbhy99}) and it can
further reduce the importance of $\eta_{ee}$ (regardless the
details of \emph{Umklapp} freezing).

The contribution of electron-electron collisions to $\eta_e$ is
not so high as their contribution to the electron thermal
conductivity $\kappa_e$ \cite{sy06}. This is because $\kappa_{ee}$
stronger depends on the screening momenta than $\eta_{ee}$. As a
result, the dynamical Landau damping introduces into $\kappa_{ee}$
an additional factor proportional to $T$. In the asymptotic region
IV, $\kappa_{ee}$ becomes then temperature-independent. In
contrast, the Landau damping introduces into $\eta_{ee}$ a much
weaker factor $\propto T^{1/3}$.

When calculating the electron-electron collision rate we have
neglected the ion contribution into the  polarization functions
$\Pi_l$ and $\Pi_t$. It is a good approximation in the case of
weak Coulomb coupling of ions $T\gtrsim Z^2e^2/(a k_B)$, when the
ions constitute a nearly ideal Boltzmann gas [$a=(4\pi
n_i/3)^{-1/3}$ being the ion-sphere radius, determined by the ion
number density $n_i$]. The calculation of the ion contribution to
the plasma screening at lower temperatures $T\lesssim Z^2e^2/(a
k_B)$ (where the ions form a Coulomb liquid or solid) is a
complicated and unsolved problem. We have also neglected the
effects of strong magnetic fields which can be available in
neutron star envelopes and which can greatly modify the electron
shear viscosity. These effects can be divided into two groups.
First, they are the classical effects of electron magnetization
owing to a fast electron rotation about magnetic field lines.
Second, there are the effects of the Landau quantization of
electron motion in a magnetic field (important usually for higher
magnetic fields than the magnetization effects). The
generalization of our solution to not too high magnetic fields,
that do not affect the polarization functions, is straightforward.
Stronger magnetic fields make the polarization tensor anisotropic,
dependent of the magnetic field strength and direction, The
effects of the ion polarization and strong magnetic fields on
$\eta_{ee}$ are beyond the scope of the present paper.

\section{Shear viscosity in the cores of pulsating
pre-white dwarf and white dwarf stars}
\label{white dwarfs}

Asteroseismology of white dwarfs is a rapidly developing
field. More than 150 pulsating white dwarfs have already been
observed (e.g., \cite{k07}, \cite{w98} and references therein).
The pulsation periods range from few minutes to few tens of minutes,
and the relative pulsation amplitudes (pulsating fraction of
star's luminosity) can be as high as a few percent.
A comparison of observed and theoretical pulsation
frequencies allows one to identify pulsation modes,
to accurately determine white dwarf masses and radii,
and to explore their rotation, magnetic fields,
internal structure and evolution.

All observed pulsation modes are interpreted as non-radial
gravity modes (g-modes, produced owing to buoyancy forces)
of multi-polarity $\ell=1,2$ with
$k=1,\ldots 50$ radial nodes. They are
excited in the envelopes of young warm  pre-white
dwarfs and white dwarfs presumably by the instability that is
mainly associated with partial
ionization of the plasma.

Pulsating pre-white dwarfs (called PG 1159, or GW Vir stars) have
the effective surface temperatures $T_\mathrm{eff}$ from $\simeq$
170~000 K to 75~000 K. They are hot, young stars (of age $\lesssim
10^5$ yr), still contracting slowly in the course of cooling
because of a not too strong electron degeneracy in their cores.
Their pulsations are thought to be driven by the partial
ionization of C and O in the respective layers (which requires
high $T_\mathrm{eff}$). In pulsating white dwarfs with helium
atmospheres (DBV, or V777 Her stars), $T_\mathrm{eff}$ ranges from
$\simeq$ 29~000 K to 22~000 K; these pulsations are excited in the
He partial ionization zone. Pulsating white dwarfs with hydrogen
atmospheres (DAV, or ZZ Ceti stars) have $T_\mathrm{eff}$ from
$\simeq$ 12~500 K to 10~500 K. Their pulsations are most probably
excited by convection in the outer layer with the partial
ionization of hydrogen.  Also, there exists a population of
pulsating white dwarfs which belong to cataclysmic variables
(accreting binaries). These white dwarfs have different
composition of surface layers (contaminated by accretion from a
companion star). As a result, they have different partial
ionization zones, and their pulsations can be excited in a wide
range of $T_\mathrm{eff}$ (from 10~000 K to 20~000 K and higher).

Let us analyze the shear viscosity in the cores of pulsating
white dwarfs. This viscosity participates in the
damping of those pulsations which
penetrate into the core. In a steadily pulsating star
(whose pulsations are generated by a driving force)
the viscosity can limit the pulsation amplitude.

\begin{figure}
\begin{center}
\includegraphics[width=1.0\textwidth]{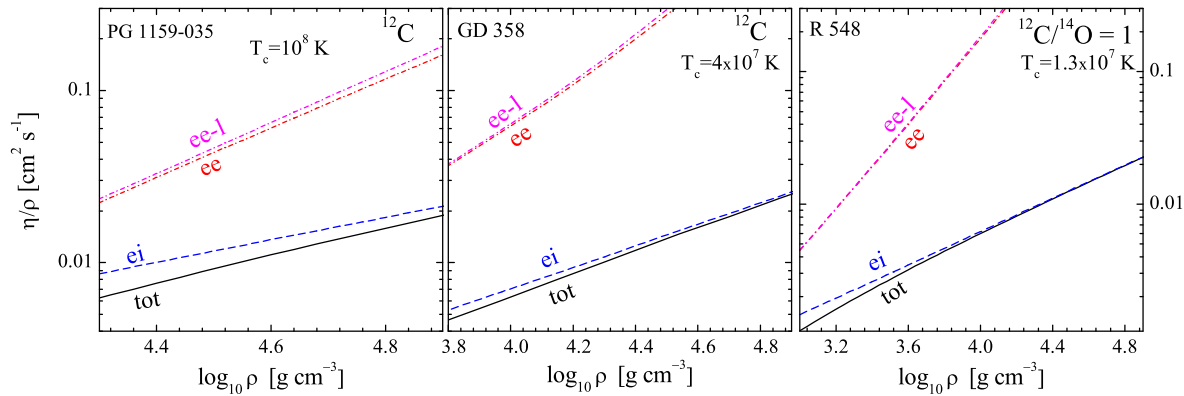}
\caption{Kinematic shear viscosity $\eta/\rho$ in the core of the
pre-white dwarf PG 1159--035 (left), the DBV white dwarf GD 358
(middle) and the DBA white dwarf R 548 (right). The curves `tot'
show the total viscosity; the curves `ei' and `ee' are the
contributions of electron-ion and electron-electron collisions,
respectively. Lines `ee-l' refer to electron-electron collisions
retaining the exchange of longitudinal plasmons alone. The
vertical scale on the right panel is different
(shown on the right axis).}
\label{F:GDPG}
\end{center}
\end{figure}

For illustration, in Figure \ref{F:GDPG} we
plot the density dependence of the
kinematic shear viscosity $\eta/\rho$
in the cores of three pulsating stars (three panels) ---
one pre-white dwarf and two white dwarfs.
A solid curve marked `tot' on each panel presents the
total electron viscosity $\eta/\rho$, while dashed and
dash-dotted curves
`ei' and `ee' are the partial contributions of
electron-ion and electron-electron collisions, respectively;
a dash-dotted line `ee-l' is the electron-electron
contribution owing to the exchange of longitudinal
plasmons alone.

The left panel of Figure \ref{F:GDPG} corresponds to the
conditions in the core of the extremely hot pulsating pre-white
dwarf PG 1159--035. Its pulsation spectrum is rich; one has
identified $\approx$200 pulsation modes \cite{cetal07}. An
analysis of astroseismological and spectral observations gives,
among other things, the effective surface temperature
$T_\mathrm{eff}\approx 140~000$~K, and the star's mass $\approx
0.6\,M_\odot$ \cite{cetal07}. According to the theory \cite{ok00},
the internal temperature of this star is mainly determined by the
neutrino emission; it is nearly constant over the core, being
close to the central temperature $T_c \approx 10^8$~K. Thus, we
have plotted the shear viscosity for the isothermal  degenerate
core assuming pure carbon composition. The highest density $\rho$
in the figure approximately corresponds to the central density of
the star. At the lowest density, the electron degeneracy becomes
mild (while our results are limited by a strong degeneracy). One
can see that the electron-electron contribution to the electron
shear viscosity is important only in the outer core, at $\rho
\lesssim 10^5$ g~cm$^{-3}$. The contribution of the
electron-electron collisions due to the exchange of transverse
phonons is small, but noticeable. It reduces
the total electron viscosity $\eta_\mathrm{e}$,
maxiumum by a factor of 1.5, and the reduction increases with
growing $\rho$.

The middle panel of Figure \ref{F:GDPG} is appropriate for the
conditions in the core of the pulsating DBV white dwarf GD~358. It
was the first DBV pulsating dwarf observed, with at least ten
well-identified pulsation modes \cite{bw94} ($\ell=1$, $k \lesssim
20$). Observations of GD~358 yield its mass $\approx 0.6~M_\odot$
and the effective surface temperature $T_\mathrm{eff}\approx
24~000$~K. These parameters and evolutionary models, reviewed in
Ref.\ \cite{tfw90}, imply the central temperature $T_c\approx
4\cdot 10^7$~K and the central density $\rho_c\approx 2\cdot
10^6$~g~cm$^{-3}$. We have evaluated the shear viscosity (Figure
\ref{F:GDPG}) using an appropriate theoretical temperature profile
in the white dwarf core from Ref.\ \cite{tfw90}. For illustration,
we have selected the pure carbon composition of the degenerate
core. One can see, that the electron-electron contribution
$\eta_\mathrm{ee}$ to the total electron shear viscosity
$\eta_\mathrm{e}$ is now less important, about $10-20\%$. The
boundary of the degenerate core shifts now to lower densities (to
$\rho \sim 3 \cdot 10^3$ g~cm$^{-3}$); the contribution of the
electron-electron collisions via the exchange of transverse
plasmons is lower.

Finally, the right panel of Figure \ref{F:GDPG} shows the shear
viscosity for the conditions in the core of the pulsating DAV
white dwarf R~548. The observational data imply $T_\mathrm{eff}
\approx$ 12~000 K, and, again, the white dwarf mass $\approx
0.6\,M_\odot$. The theory \cite{tfw90} predicts a nearly
isothermal degenerate core with the temperature $T \approx T_c
\approx 1.3\cdot10^7$ K. We have calculated the shear viscosity
assuming the C-O core with equal mass fractions of C and O. The
shear viscosity $\eta_{ei}$, limited by electron-ion collisions in
a C-O mixture, has been determined using the linear mixing rule
\cite{hpy07}. The contribution of collisions via the exchange of
transverse plasmons is now noticeable only at $\rho\lesssim 4\cdot
10^3$~g~cm$^{-3}$.

Our three examples in Figure \ref{F:GDPG} cover a representative
range of pulsating pre-white dwarf and white dwarf models. In all
the cases the kinematic electron shear viscosity $\eta/\rho$ is of
the same order of magnitude, with characteristic values of
$\eta/\rho \sim 10^{-3}\ -\ 10^{-2}$ cm$^2$~s$^{-1}$. If the
damping of g-modes were solely determined by shear viscosity in
the stellar core, a typical damping time could have been estimated
as $\tau_\mathrm{shear}\sim \lambda^2 \rho /\eta$, where $\lambda$
is a length scale of pulsation modes. Characteristic length scales
are $\lambda \sim R/k \sim(0.1-0.01)R$, where $R \sim$ 10~000 km
is the white dwarf radius and $k$ is the number of radial
oscillation nodes. This gives $\tau_\mathrm{shear} \sim
10^8-10^{11}$ yr, indicating that the shear viscosity in the cores
of pulsating white dwarfs is rather inefficient in damping
observed pulsations. It would be equally inefficient to damp other
possible large-scale hydrodynamic motions in white dwarf cores
(for instance, differential rotation).

The shear viscosity in white dwarf cores should have much stronger
effect on small-scale motions. For instance, these could be
small-scale (high-$k$ or high-$\ell$) pulsations, or ordinary
pulsations in stratified cores, containing sharp boundaries
between different layers (e.g., separating phases of
different elements or phases of solidified and liquid matter).
Viscous dissipation in boundary layers can be strong.

Similar conclusions can be made on the efficiency of the electron
shear viscosity in the envelope (crust) of neutron stars (where
one typically has $\eta_e/\rho \sim$ 0.01--100 cm$^2$ s$^{-1}$ for
the range of the internal crust temperature from $\sim 10^9$~K to
$10^7$~K \cite{chy05}). The viscosity $\eta_e$ there is not too
high and cannot be a strong regulator of large-scale motions
(pulsations). Nevertheless, characteristic length scales in
neutron stars are naturally much shorter than in white dwarfs
(because neutron stars are smaller, and their crust thickness is
$\lesssim$1 km). This increases the efficiency of the shear
viscosity (decreases viscous damping times) in neutron star
envelopes (as compared to white dwarf cores). Moreover, a neutron
star crust has a well defined heterogeneous (multi-layer)
structure \cite{hpy07}, where viscous boundary layers can occur
and viscous dissipation can be strong. Such effects are currently
almost unexplored.


\section{Conclusion}
\label{conclus}

We have calculated the partial
electron shear viscosity $\eta_{ee}$ owing to
collisions between degenerate electrons in a dense degenerate
plasma taking into account the Landau damping of transverse
plasmons. Our main conclusions are:
\begin{enumerate}

\item{} The Landau damping reduces $\eta_{ee}$
for all temperature and density
regions I--IV (Table \ref{tab:regimes}).

\item{} The strongest reduction occurs in the region IV of cold
($T\lesssim T_{pe}$) relativistic ($\rho\gg 10^{6}$~g~cm$^{-3}$)
electron plasma. In this region, the Landau damping lowers the
viscosity $\eta_{ee}$ by several orders of magnitude and modifies
its temperature dependence (which becomes $\eta_{ee}\propto
T^{-5/3}$ instead of the traditional dependence $\eta_{ee}\propto
T^{-2}$).

\item{} The viscosity $\eta_{ee}$ gives a noticeable
contribution to the total electron viscosity $\eta_e$
in a plasma of light ions at $T \gtrsim T_{pe}$.

\item{} The viscosity $\eta_{ee}$ can also give
a significant contribution to $\eta_e$ in a deep
crust ($\rho \gtrsim 10^{13}$ g~cm$^{-3}$) of a
cold neutron star ($T \lesssim 10^{7}$) provided
crystalline lattice of atomic nuclei is rather
pure and the freezing of \emph{Umklapp} processes
at low $T$
is sufficiently strong. Both factors,
the impurity of crystals and the character of
\emph{Umklapp} freezing, are currently uncertain
and require special study.

\end{enumerate}

We have approximated our results by simple analytic expressions
which are valid in the wide range of densities and temperatures
appropriate to degenerate cores of white dwarfs and red giants and
to the envelopes of neutron stars. The analytic approximations can
be easily incorporated into computer codes to simulate
hydrodynamical processes in these stars. We have briefly analyzed
(Section \ref{white dwarfs}) the efficiency of viscous damping in
the cores of white dwarfs and envelopes of neutron stars.

The contribution of collisions of degenerate electrons via the
Landau damping during the
exchange of transverse plasmons has been neglected in all previous
considerations of the electron shear viscosity. In the context of transport
properties of dense quark plasma it was studied by Heiselberg and
Pethick \cite{hp93}. We have recently reconsidered the effect of
the Landau damping on
the electron thermal conductivity in neutron star crust and core
\cite{sy06,sy07}. The same effect on
the shear viscosity in the neutron star core (taking into account
possible superfluidity of nucleons)
will be studied in a similar way and published elsewhere.

\ack{ I am grateful to A.~I.~Chugunov for sharing with me
unpublished results and for noting the mislead in the preliminary
version of the paper. This work was partially supported by the
Dynasty Foundation, by the Russian Foundation for Basic Research
(grants 05-02-16245, 05-02-22003), and by the Federal Agency for
Science and Innovations (grant NSh 9879.2006.2).}

\end{document}